\documentstyle[12pt,epsf,epsfig]{article}

\newcommand{\be}{\begin{equation}}
\newcommand{\ee}{\end{equation}}
\newcommand{\bea}{\begin{eqnarray}}
\newcommand{\eea}{\end{eqnarray}}

\def\gappeq{\mathrel{\rlap {\raise.5ex\hbox{$>$}}
{\lower.5ex\hbox{$\sim$}}}}

\begin{document}
\begin{titlepage} 

\begin{flushright} 
ACT-07/00 \\ 
CERN--TH/2000--155 \\
CTP-TAMU-17/00 \\
gr-qc/0006004
\end{flushright} 

\begin{centering} 

{\large {\bf Dynamical Formation of Horizons in Recoiling \\
$D$ Branes}}

\vspace{0.5in}

{\bf John Ellis}~$^{a}$, {\bf N.E. Mavromatos}~$^{b}$
and {\bf D.V. Nanopoulos}~$^{c,d,e}$

\end{centering} 

\vspace{0.2in} 

\noindent $^a${\it Theory Division, CERN, CH-1211 Geneva 23, Switzerland}\\
$^b$ {\it Theoretical Physics Group, Department of Physics,
King's College London, Strand, London WC2R 2LS, UK}\\
$^c$ {\it Department of Physics, Texas A \& M University, 
College Station, TX~77843-4242, USA},\\
$^d$ {\it Astroparticle Physics Group, Houston
Advanced Research Center (HARC), 
Mitchell Campus,
Woodlands, TX 77381, USA,}\\
$^e$ {\it Academy of Athens, Chair of Theoretical Physics, 
Division of Natural Sciences, 
28~Panepistimiou Avenue, 
Athens 10679, Greece}

\begin{center}

{\bf Abstract}

\end{center}

{\small A toy calculation of string/$D$-particle interactions within a
world-sheet approach indicates that quantum recoil effects - reflecting
the gravitational back-reaction on space-time foam due to the propagation
of energetic particles - induces the appearance of a microscopic event
horizon, or `bubble', inside which stable matter can exist. The scattering
event causes this horizon to expand, but we expect quantum effects to
cause it to contract again, in a `bounce' solution. Within such `bubbles',
massless matter propagates with an effective velocity that is less than
the velocity of light {\it in vacuo}, which may lead to observable
violations of Lorentz symmetry that may be tested experimentally.  The
conformal invariance conditions in the interior geometry of the bubbles
select preferentially three for the number of the spatial dimensions,
corresponding to a consistent formulation of the interaction of $D3$
branes with recoiling $D$ particles, which are allowed to fluctuate
independently only on the $D3$-brane hypersurface.}

\end{titlepage} 

\section{Introduction}

The discovery of $D$ branes~\cite{Polch} has revolutionized the study of
black-hole physics. Now one has quasi-realistic string models
of black holes in different dimensions, which one can use to study
profound issues concerning the reconciliation of general relativity
and quantum mechanics. A key breakthough was the
demonstration that the entropy of a stringy black hole
corresponds to the number of its distinct quantum states~\cite{Whair,Sen}.
Thus $D$ branes offer the prospect of accounting exactly for the
flow of information in processes involving particles and
black holes. However, it is not immediately apparent whether an
observer will perceive information to be lost in any given
particle/$D$-brane interaction: the answer depends whether
she is able to recover all the information transferred from the
scattering particle to the recoiling black hole.
It is important to address this issue at both the macroscopic
and microscopic levels, where the answers may differ. In the case
of a macroscopic black hole, it is difficult to see how {\it in
practice} all the quantum information may be recovered 
without a complete set of observations of the emitted Hawking
radiation~\cite{Hawking}. However, even if this is possible {\it in
principle},
the problem of the microscopic `end-game' that terminates the
Hawking evaporation process is unsolved, in our view.

It may be useful to recall one of the intuitive ways of
formulating the information loss in the process of Hawking radiation
from a macroscopic black hole, whose stringy analogue we study in
this paper. Consider the quantum-mechanical creation of a pure-state 
particle pair $\vert A,B \rangle$
close to the (classical) black-hole horizon of such a macroscopic black
hole. One can then envisage that particle $B$ falls inside this
horizon, whilst particle $A$ escapes as Hawking radiation. The quantum
state of the
particle $B$ is apparently unobservable, and hence information is
apparently lost. One may represent the corresponding quantum process as
\begin{equation}
\vert A, B \rangle \; + \; \vert BH \rangle \rightarrow
\vert BH + B_i \rangle \; + \; \vert A_i \rangle.
\label{intuition}
\end{equation}
The observable subset $A_i$ of the final state can only be
represented by a density matrix
\begin{equation}
\rho_A \; \equiv \; \Sigma_i \vert A_i \rangle \langle A_i \vert,
\label{mixedrho}
\end{equation}
and it appears that the pure state $\vert A, B \rangle$ evolves
into a mixed density matrix $\rho_A$, representing (almost) a thermal
state.

This argument is very naive, and one would like to formulate a
more precise treatment of this process at the microscopic level,
suitable for describing space-time foam~\cite{foam}.
The purpose of this paper is to take a step towards this goal using
a stringy treatment~\cite{kogan96,ellis96,mavro+szabo} of the interaction
between
closed-string particle `probes' and $D$-brane black holes.
We have developed previously an approach capable of accommodating the
recoil of a $D$-brane black hole struck by a closed-string `probe',
including also quantum effects associated with higher-genus
contributions to the string path integral. We have shown
explicitly~\cite{kanti98,ellis96} how the loss of information to the
recoiling
$D$ brane (assuming that it is unobserved) leads to information loss,
for both the scattered particle and also any spectator particle. This
information loss can be related to a change in the background metric
following the scattering event, which can be regarded as creating an
Unruh-like `thermal' state.

In this paper, we take this line of argument a step further, by
demonstrating that closed-string particle/$D$-brane scattering
leads in general to the formation of a microscopic event horizon,
within which string particles may be trapped. The scattering
event causes expansion of this horizon, which is eventually halted
and reversed by Hawking radiation~\cite{Hawking}. Thus we have a
microscopic
stringy realization of the process (\ref{mixedrho}) discussed
intuitively above.

A peculiarity of this approach is that the conformal invariance conditions
select preferentially backgrounds with three spatial dimensions. This
leads to a consistent formulation of the interaction of $D3$ branes with
recoiling $D$ particles, which are allowed to fluctuate independently only
on the $D3$-brane hypersurface. 

\section{Formulation of $D$-Brane Recoil}

As discussed in references~\cite{kogan96,ellis96,mavro+szabo}, the
recoil of a $D$-brane string soliton after
interaction with a closed-string state is
characterized
by a $\sigma$ model on the string world sheet $\Sigma$, that is deformed
by a pair of logarithmic operators~\cite{lcft}:
\begin{equation}
C^I_\epsilon = \epsilon \Theta_\epsilon (X^I),\qquad
D^I_\epsilon = X^I \Theta_\epsilon (X^I), \qquad I \in \{0,\dots, 3\}
\label{logpair}
\end{equation}
defined on the boundary $\partial \Sigma$ of the string world
sheet. Here $X^I, I\in \{0, \dots, p\}$ obey Neumann boundary
conditions on $\Sigma$, and denote the $D$-brane
coordinates, whilst $\epsilon\rightarrow0^+$ is
a regulating parameter and $\Theta_{\epsilon}(X^{I})$ is a
regularized Heaviside step function.
The remaining $y^i, i\in \{p+1, \dots, 9\}$ in (\ref{logpair}) denote the
transverse bulk directions. For reasons of convergence of the
world-sheet path integral,
we take the space-time $\{ X^I, y^i \}$ to have {\it Euclidean} 
signature.

In the case of $D$ particles~\cite{kogan96,ellis96,mavro+szabo},  
the index $I$ takes the
value $0$ only, in which case the operators (\ref{logpair}) act as
deformations of the conformal field theory on the world sheet. The
operator
\begin{equation}
u_i \int _{\partial \Sigma} \partial_n X^i D_\epsilon
\label{movement}
\end{equation}
describes the movement of the $D$ brane induced by the scattering,
where $u_i$ is its recoil velocity, and $y_i \int _{\partial
\Sigma} \partial_n X^i C_\epsilon $ describes quantum fluctuations
in the initial position $y_i$ of the $D$ particle. It has been
shown rigorously~\cite{mavro+szabo} that the logarithmic conformal
algebra ensures energy--momentum  conservation during the recoil
process:
\begin{equation}
u_i = ( k^1_i + k^2_i)/M_D~, 
\label{conservation}
\end{equation}
where $k^1 (k^2)$ is
the momentum of the propagating closed string state before (after)
the recoil, and $M_D=1/(\ell _s g_s)$ is the mass of the $D$ brane,
where $g_s$ is the string coupling, which is assumed
here to be weak enough to ensure that the $D$ brane is very massive,
and $\ell _s$ is the string
length.

The second member of the logarithmic pair of $\sigma$-model 
deformations is
\begin{equation}
y_{i} \int _{\partial \Sigma} \partial_n X^i C_\epsilon~,
\label{logpair2}
\end{equation}
where, in order to realize the logarithmic
algebra between the operators $C$ and $D$, one uses
as a regulating parameter~\cite{kogan96}
\begin{equation}
\epsilon^{-2} \sim \ln [L/a] \equiv \Lambda,
\label{defeps}
\end{equation}
where $L$ ($a$) is an infrared (ultraviolet) world--sheet cutoff.
The recoil operators (\ref{logpair2}) are
relevant, in the sense of the renormalization group for the
world--sheet field theory, having small conformal dimensions
$\Delta _\epsilon = -\epsilon^2/2$. Thus the $\sigma$-model
perturbed by these operators is not conformal for $\epsilon \ne
0$, and the theory requires Liouville
dressing~\cite{david88,distler89,ellis96}. The consistency of this
approach is supported by the above-mentioned proof of momentum
conservation during the scattering process~\cite{mavro+szabo}.

As discussed in~\cite{ellis96,kanti98}, the deformations
(\ref{logpair}) create a local distortion of the space-time
surrounding the recoiling $D$ brane, which may also be
determined using the method of Liouville dressing.
In~\cite{ellis96,kanti98} we concentrated on describing the
resulting space-time in the case when a $D$-particle defect embedded in
a $D$-dimensional space-time recoils after the scattering of a
closed string. To leading order in
the recoil velocity $u_i$ of the $D$ particle, the resulting
space-time was found, for times $t \gg 0$ long after the
scattering event at $t=0$, to be equivalent to a Rindler wedge,
with apparent `acceleration' $\epsilon u_i$~\cite{kanti98}, where
$\epsilon$ is defined above (\ref{defeps}).
For times $t < 0$, the space-time is flat Minkowski~\footnote{There is
hence a discontinuity at $t =0$, which leads to particle
production and decoherence for a low-energy spectator field theory
observer who performs local scattering experiments
long after the scattering, and far away from the
location of the collision of the closed string with the
$D$ particle~\cite{kanti98}.}.

This situation is easily generalized to $Dp$ branes~\cite{emw99}.
The folding/recoil deformations of the $Dp$ brane (\ref{logpair2})
are relevant deformations, with anomalous dimension
$-\epsilon^2/2 $, which disturbs the conformal invariance of the
world-sheet $\sigma$ model, and restoration of conformal invariance
again requires
Liouville dressing~\cite{david88,distler89,ellis96}, as discussed above.
To determine the effect of
such dressing on the space-time geometry, it is essential  to
write~\cite{ellis96} the boundary recoil deformations as  bulk
world-sheet deformations
\begin{equation}
\int _{\partial \Sigma} {\overline g}_{Iz} x \Theta_\epsilon (x)
\partial_n z =
\int _\Sigma \partial_\alpha \left({\overline g}_{Iz} x \Theta_\epsilon
(x)
\partial ^\alpha z \right)
\label{a1}
\end{equation}
where the ${\overline g}_{Iz}$ denote renormalized
folding/recoil couplings~\cite{mavro+szabo}. Such
couplings are marginal on a flat world sheet, and
the operators (\ref{a1}) are marginal also on a curved
world sheet, provided~\cite{distler89} one dresses the (bulk)
integrand by multiplying it by a factor $e^{\alpha_{Ii}\phi}$,
where $\phi$ is the Liouville field and $\alpha_{Ii}$ is the
gravitational conformal dimension. This is related to the
flat-world-sheet anomalous dimension $-\epsilon^2/2$ of the recoil
operator, viewed as a bulk world-sheet deformation
by~\cite{distler89}:
\begin{equation}
\alpha_{Ii}=-\frac{Q_b}{2} +
\sqrt{\frac {Q_b^2}{4} + \frac {\epsilon^2}{2} }
\label{anom}
\end{equation}
where $Q_b$ is the central-charge deficit of the bulk world-sheet
theory. In the recoil problem at hand, as discussed
in~\cite{kanti98},
\be
Q_b^2 \sim \epsilon^4/g_s^2  > 0
\label{centralcharge}
\ee
for weak folding deformations $g_{Ii}$, and hence one is
confronted with a {\it supercritical} Liouville theory. This
implies a {\it Minkowskian-signature} Liouville-field kinetic term
in the respective $\sigma$ model~\cite{aben89}, which prompts one
to interpret the Liouville field as a time-like target
field. 

There are two approaches which one can follow at this point.
In the first of them~\cite{leonta}, 
this time is considered as a {\it second} time
coordinate~\cite{emn98}, which is independent of the
(Euclideanized) $X^0$. The presence of this second `time'
does not affect physical observables, which are defined
for appropriate slices with fixed Liouville coordinate, e.g., $\phi
\rightarrow \infty$ or equivalently
$\epsilon \rightarrow 0$.
From the expression (\ref{centralcharge}) we conclude (cf.
(\ref{anom})) that $\alpha_{Ii} \sim \epsilon $ to leading order
in perturbation theory in $\epsilon$, to which we restrict
ourselves here. 
In the second approach~\cite{emn98}, which we shall 
mainly follow here, the (Minkowskian) Liouville 
field $\phi$ is identified with the (initially Euclidean) coordinate
$X^0$, and hence one is no longer considering constant Liouville
field slices. In this approach, however, one still identifies 
$\epsilon^{-2}$ with the target time, which in turn implies that 
the perturbative world-sheet approach is valid,
provided one works with sufficiently large times $t$,
i.e. small $\epsilon^2$.

We next remark~\cite{ellis96} that
the $X^I$-dependent field operators
$\Theta_\epsilon (X^I)$ scale as follows with $\epsilon$:
$\Theta_\epsilon(X^I) \sim e^{-\epsilon X^I}
\Theta(X^I)$, where $\Theta(X^I)$ is a Heavyside step function
without any field content, evaluated in the limit $\epsilon \rightarrow 0^+$.
The bulk deformations, therefore, yield the following
$\sigma$-model terms:
\begin{equation}
\frac{1}{4\pi \ell_s^2}~\int _\Sigma 
\sum_{I=0}^{3} \left( \epsilon^2 {\overline g}^C_{Ii} + \epsilon 
{\overline g}_{Ii} X^I\right)
e^{\epsilon(\phi_{(0)} - X^I_{(0)})}\Theta(X^I_{(0)})
\partial_\alpha \phi 
\partial^\alpha y_i~
\label{bulksigma}
\end{equation}
where the subscripts $(0)$ denote world-sheet zero modes, and 
${\overline g}^C_{0i}=y_i$.

Upon the interpretation of the Liouville zero mode $\phi_{(0)}$ as
a (second)
time-like coordinate, the deformations (\ref{bulksigma}) yield
metric
deformations of the generalized space-time
with two times. The metric components
for fixed Liouville-time slices can be
interpreted~\cite{ellis96}
as expressing the distortion of the space-time
surrounding the recoiling $D$-brane soliton.

For clarity,
we now drop the subscripts $(0)$ for the rest of this paper,
and we work in a region of space-time
such that $\epsilon (\phi - X^I)$ is finite
in the limit $\epsilon \rightarrow 0^+$.
The resulting space-time distortion is therefore
described by the metric elements
\begin{eqnarray}
&~& G_{\phi\phi} = -1, \qquad G_{ij} =\delta_{ij}, \qquad
G_{IJ}=\delta_{IJ}, \qquad G_{iI}=0,   \nonumber \\
&~& G_{\phi i} = \left(\epsilon^2 {\overline g}^C_{Ii} +
 \epsilon {\overline g}_{Ii}X^I \right)\Theta (X^I)~,
\qquad i=4, \dots 9,~~I=0, \dots 3
\label{gemetric}
\end{eqnarray}
where the index $\phi $ denotes Liouville `time', not to be confused
with the Euclideanized time which is one of the $X^I$.
To leading order in $\epsilon {\overline g}_{Ii}$,
we may ignore the $\epsilon^2 {\overline g}^C_{Ii}$ term.
The presence of $\Theta(X^I)$ functions and
the fact that we are working in the region $y_i >0$
indicate that
the induced space-time is piecewise continuous~\footnote{The
important implications for non-thermal particle production
and decoherence for a spectator low-energy field theory
in such space-times were discussed in~\cite{kanti98,ellis96}, where
the $D$-particle recoil case was considered.}.
In the general recoil/folding case considered in this article,
the form of the resulting patch of the surrounding
space-time can be determined fully if one computes
the associated curvature tensors, along the lines
of~\cite{kanti98}.

We next study in more detail some physical aspects of
the metric (\ref{gemetric}),
restricting ourselves, for simplicity, to the case
of a single Dirichlet dimension $z$ that
plays the r\^ole of a bulk dimension
in a set up where there are 
Neumann coordinates $X^I$, $I=0,\dots3$
parametrizing a D4 (Euclidean) brane, interpreted as
our four-dimensional space-time.
Upon performing the time transformation
$\phi \rightarrow \phi - \frac{1}{2}\epsilon {\overline g}_{Iz} X^I z $, the
line element (\ref{gemetric}) becomes:
\begin{eqnarray}
&~&ds^2 =-d\phi^2 + \left(\delta_{IJ}
-\frac{1}{4}\epsilon^2{\overline g}_{Iz}{\overline g}_{Jz}~z^2\right)~dX^I dX^J
+  
\nonumber \\
&~& \left(1 + \frac{1}{4}\epsilon ^2
{\overline g}_{Iz}{\overline g}_{Jz}~X^I~X^J\right)~dz^2 -
\epsilon {\overline g}_{Iz}~z~dX^I~d\phi~, \nonumber \\
\label{bendinglineel}
\end{eqnarray}
where $\phi$ is the Liouville field which, we remind the reader,
has Minkowskian signature in the case of supercritical
strings that we are dealing with here.

One may now make a general coordinate transformation on the
brane $X^I$ that diagonalizes the pertinent induced-metric
elements in (\ref{bendinglineel})~\footnote{Note that general
coordinate invariance is assumed to be a good symmetry on the
brane, away from the `boundary' $X^I=0$.}. For instance, to
leading order in the deformation couplings ${\overline
g}_{Iz}{\overline g}_{Jz}$, one may redefine the $X^I$ coordinates
by 
\begin{eqnarray} X^I &\rightarrow& X^I -\frac{\epsilon^2}{8}z^2
{\overline g}_{Iz} \sum_{J \ne I}{\overline g}_{Jz}X^J,\nonumber\\
z &\rightarrow& z \left(1 + \frac{\epsilon^2}{8}\sum_{I \ne J}
{\overline g}_{Iz}{\overline g}_{Jz}X^{I}X^{J}\right)
\end{eqnarray}
which leaves only diagonal elements  of the
metric tensor on the (redefined) hyperplane $X^I$. In this case,
the metric becomes, to leading order in $g_{Iz}^2$ and
in the case where $
\epsilon {\overline g}_{Iz}z << 1$:
\begin{eqnarray}
&~&ds^2 =-d\phi^2 + \left(1
-\alpha^2 ~z^2\right)~(dX^I)^2
+ \left(1 + \alpha^2 ~(X^I)^2\right)~dz^2 - \epsilon {\overline g}_{Iz}~z~dX^I~d\phi~,
\nonumber \\
&~& \alpha=\frac{1}{2}\epsilon {\overline g}_{Iz} \sim g_s |\Delta P_z|/M_s
\label{bendinglineel3}
\end{eqnarray}
where the last expression is a reminder that
one can express the parameter
$\alpha$ (in the limit $\epsilon \rightarrow 0^+$)
in terms of the (recoil) momentum transfer $\Delta P_z$ along the bulk
direction.

A last comment, which is important for our purposes here, 
concerns the case in which
the metric (\ref{bendinglineel3}) is {\it exact}, i.e., it holds
to all orders in ${\overline g}_{Iz}z$.
This is the case where
there is no world-sheet
tree-level momentum transfer. This naively corresponds to the case
of static intersecting branes. However, the whole philosophy of
recoil~\cite{kogan96,mavro+szabo} implies that, even in that case,
there are quantum fluctuations induced by the sum over genera of the
world sheet. The latter implies the existence of a statistical
distribution of logarithmic deformation couplings of Gaussian type
about a mean-field value ${\overline g}^{C}_{Iz}=0$. Physically,
the couplings
${\overline g}_{Iz}$ represent recoil velocities of the intersecting
branes,
hence these Gaussian fluctuations
represent the effects of quantum fluctuations about the
zero recoil-velocity case, which may be considered as quantum
corrections to the static intersecting-brane case.
We therefore consider a statistical average
$<< \cdots >>$ of the line element (\ref{bendinglineel})
\begin{eqnarray}
&~&<<ds^2>> =-d\phi^2 + \left(1
-\frac{1}{4}\epsilon^2
<<{\overline g}_{Iz}{\overline g}_{Jz}>>~z^2\right)~dX^I dX^J
+ \nonumber \\
&~& \left(1 + \frac{1}{4}\epsilon ^2
 <<{\overline g}_{Iz}{\overline g}_{Jz}>>~X^I~X^J\right)~dz^2
- \epsilon <<{\overline g}_{Iz}>>~z~dX^I~d\phi~, \nonumber \\
\label{bendinglineel2}
\end{eqnarray}
where
\be
<< \cdots >>=\int _{-\infty}^{+\infty}d{\overline g}_{Iz}
\left(\sqrt{\pi}\Gamma \right)^{-1} 
 e^{-{\overline g}_{Iz}^2/\Gamma^2} (\cdots) \label{gauss}
\ee
and the width $\Gamma$ has been calculated
in \cite{mavro+szabo}, 
is found after summation over world-sheet
genera to be
proportional to the string coupling $g_s$. 
In fact, it can be shown~\cite{mavro+szabo} that $\Gamma$ scales 
as $\epsilon {\overline \Gamma}$, where ${\overline \Gamma}$ 
is independent of $\epsilon$. This will be important later on, when 
we consider the identification of $\epsilon$ with the target time $t$.

We see from (\ref{gauss}),
assuming that $g_{Iz}=|u_i|$ where $u_i=g_s \Delta P_i/M_s$
is the recoil velocity~\cite{kogan96,mavro+szabo}, that
the average line element
$ds^2$ becomes:
\begin{eqnarray}
&~&<<ds^2>> =-d\phi^2 + \left(1
-\alpha^2 ~z^2\right)~(dX^I)^2
+ \left(1 + \alpha^2 ~(X^I)^2\right)~dz^2,
\nonumber \\
&~& \alpha=\frac{1}{2\sqrt{2}}\epsilon^2 {\overline \Gamma}
\label{bendinglineel3ab}
\end{eqnarray}
The definition of $\alpha$ comes from evaluating the quantity
$<<{\overline g}_{Iz}^2>>$ using the statistical distribution (\ref{gauss}).
Thus the average over quantum fluctuations leads to a
metric of the form (\ref{bendinglineel3}), but with a parameter
$\alpha$ determined by the width (uncertainty)
of the pertinent quantum fluctuations~\cite{mavro+szabo}.
The metric (\ref{bendinglineel3ab})
is exact, in contrast to the metric (\ref{bendinglineel3})
which was derived for $z << 1/\alpha$. However, for our purposes
below we shall treat both
metrics as exact solutions of some string theory
associated with the recoil~\cite{leonta,emn98}.

An important feature of the line element (\ref{bendinglineel3ab}) is
the existence of a {\it horizon} at $z=1/\alpha$ for {\it Euclidean}
Neumann coordinates $X^I$. Since the Liouville field
$\phi$ has decoupled after the averaging procedure, 
one may consider slices of this field, defined by $\phi$ = const, on
which  the physics of the observable world can be studied~\cite{leonta}. 
From a
world-sheet renormalization-group view point this slicing
procedure corresponds to selecting a specific point in the
non-critical-string theory space. Usually, the infrared fixed
point
$\phi \rightarrow \infty$ is selected. In that case
one considers (\ref{defeps}) a slice for which $\epsilon^2 \rightarrow 0$.
But any other choice could do, so $\alpha$ may be considered
a small but arbitrary parameter of our effective theory.

The presence of a horizon raises the issue of how one
could analytically continue so as to pass to the space beyond the horizon.
The simplest way, compatible~\cite{leonta} with the low-energy
Einstein equations, is to take the absolute value of $1-\alpha^2 z^2$
in the metric element (\ref{bendinglineel3}) and/or (\ref{bendinglineel3ab}).
However, we prefer the second approach~\cite{emn98},
in which one identifies the (zero mode of the) Liouville mode $\phi$ with 
the time coordinate $X^0$ on the initial $Dp$ brane. 
In this case, as we shall see, the situation becomes much more 
interesting, at least  
in certain regions of the bulk space-time,
where one can calculate reliably in a world-sheet
perturbative approach. Indeed,
far away from the horizon at $|z|=1/\alpha$, 
i.e., for $\alpha ^2 z^2 << 1$, 
the line element corresponding to the space-time 
(\ref{bendinglineel3ab}) becomes, after the identification 
$\phi =X^0$:
\be
ds^2 \simeq 
-\alpha ^2 z^2 \left(dX^0\right)^2 + dz^2 + \sum_{i=1}^{3}\left(dX^i\right)^2 
\label{conical}
\ee
implying that $X^0$ plays now the r\^ole of a {\it Minkowskian}-signature
temporal variable, despite its original Euclidean nature. 
This is a result of the identification $\phi=X^0$, and the fact that 
$\phi$ appeared with Minkowskian signature due to the supercriticality 
(\ref{centralcharge}) of the Liouville string under consideration. 

Notice that although the space-time (\ref{conical}) 
is flat asymptotically as one would expect, 
and hence satisfies
Einstein's equations formally, nevertheless it
has a {\it conical} singularity  
when one compactifies the time variable $X^0$ on a circle of finite radius
corresponding to 
an inverse `temperature' $\beta$. Formally, 
this requires a Wick rotation 
$X^0 \rightarrow iX^0$ and then compactification, $iX^0=\beta e^{i\theta}$,
$\theta \in \left(0 , 2\pi \right]$.
The space-time then becomes a {\it conical} space-time of Rindler type
\be
  ds_{conical}^2 = \frac{1}{4\pi^2}\alpha^2 \beta^2 z^2 \left( d\theta \right)^2 + dz^2 
+ \sum_{i=1}^{3}\left(dX^i\right)^2 
\label{conical2}
\ee 
with deficit angle $\delta \equiv 2\pi - \alpha \beta$. 
We recall that there is a 
`thermalization theorem' for this space-time~\cite{unruh}, in the 
sense that the deficit 
disappears and the space-time becomes regular, when the temperature 
is fixed to be 
\be 
     T = \alpha /2\pi  
\label{unruh}
\ee
The result (\ref{unruh}) may be understood physically 
by the fact that $\alpha$ is essentially related to recoil. As discussed
in \cite{kanti98}, the problem 
of considering a suddenly fluctuating (or recoiling) brane at $X^0=0$,
as in our case above, 
becomes equivalent to that of an observer in a (non-uniformly)
accelerated frame. At times long after the collision the acceleration 
becomes uniform and equals $\alpha$. This implies the appearance of a
non-trivial  
vacuum~\cite{unruh}, characterized by thermal properties of the form 
(\ref{unruh}). At such a temperature the vacuum becomes 
just the Minkowski vacuum, whilst the Unruh vacuum~\cite{unruh} 
corresponds to $\beta \rightarrow \infty$. 
Here we have derived this result in a
different way than in \cite{kanti98}, but the essential physics is 
the same.

\section{$D$-Particle Recoil and the Dimensionality of the
Brane World}

\begin{figure}[htb]
\begin{center}
\epsfxsize=1in
\bigskip
\centerline{\epsffile{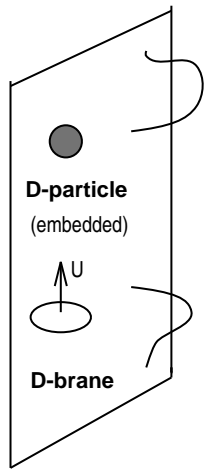}}
\caption{\it The world as a $D3$ brane 
`punctured' by $D$ particles ($D0$ branes). 
The scattering on the $D0$ brane of string states, either
closed (gravitons) or open (matter fields) that
live on the $D3$ brane, cause the $D0$ brane
to recoil, leading to stochastic effects in the
propagation of the low-energy states, as well as to 
non-zero `vacuum' energy on the $D3$ brane.} 
\label{fig3}
\end{center}
\end{figure}

In the picture envisaged above, where our world is viewed as a 
fluctuating $D$--brane, one may consider more complicated configurations
of intersecting branes. The simplest of all cases is the one
depicted in Fig.~\ref{fig3}, in which a $D$ particle is embedded in
a Euclidean $D4$ brane, which is itself embedded in a higher-dimensional
(bulk) space-time.

In this case, any low-energy string state living on the $D3$ brane
which scatters off the emebedded $D0$ brane will cause
recoil of the latter and hence distortion of space-time, according to 
the above discussion. The distortion is such as to induce
non-trivial contributions to the 
vacuum energy on the $D3$ brane, as discussed in 
detail in~\cite{emncosmol,ellis98}. 
To see this, we recall that the four-dimensional space-time,
in which the defect is embedded, is to be viewed as a bulk space-time
from the point of view of the world-sheet approach to the recoil of 
the $D$ particle. Following the same approach as that leading to 
(\ref{conical}), involving the identification of the 
Liouville field with the target time, $t$, 
one observes again that there exists an (expanding) horizon, located at 
$r^2 \equiv x_1^2 + x_2^2 + x_3^2 =t^2 /b'^2 $
where $\{ x_i \}, i=1,\dots 3$
constitute the bulk dimentions, obeying Dirichlet boundary conditions
on the world sheet, 
and $b'$ is related to the momentum uncertainty of the fluctuating 
$D$ particle. The variance $b'$ was computed~\cite{mavro+szabo}  
using a world-sheet formalism resummed over pinched 
annuli, which has been argued to be the leading-order 
effect for weak string coupling $g_s$:
\begin{equation}
(b')^2 = 4\frac{g_s^2}{\ell_s^2}\left(1 - \frac{285}{18}g_s^2 
\frac{E_{kin}}{M_Dc^2}\right)
+ {\cal O}(g_s^6) 
\label{varianceb'}
\end{equation}
where $E_{kin}$ is the kinetic energy scale of the fluctuating (heavy)
$D$ particle, $M_D=g_s/\ell_s$ is the $D$--particle mass scale, 
and $\ell_s$ is the string length. Note
the dependence of the variance $b'$ 
on the string coupling $g_s$, which arises because
quantum corrections come from the summation over world-sheet
topologies~\cite{emn98,mavro+szabo}, 
and $g_s$ is a string-loop 
counting parameter.

For the region of space-time
{\it inside the horizon} one obtains the following  metric 
on the $D3$ brane, as a result of recoil of the $D$ particle
embedded in it:
\be
ds^{2{(4)}} \simeq 
\frac{b'^2 r^2}{t^2} \left(dt\right)^2 -  
\sum_{i=1}^{3}\left(dx^i\right)^2~,\qquad r^2 = \sum_{i=1}^3 x_i^2  
< t^2/b'^2
\label{fourdime2}
\ee
Note that the scalar curvature corresponding to the metric 
(\ref{fourdime2}) 
has the form $R=-4/r^2$, and as such has a singularity 
at the initial 
location 
$r=0$ of the $D$-particle defect, as expected.

We can now check whether this metric is a solution of 
Einstein's  equations in a four-dimensional space-time $\{x_i, t \}$, 
which in our metric and signature conventions the Einstein's equations 
read 
\begin{equation}
{\cal E}_{\mu\nu}=-T_{\mu\nu}
\label{einsteinddime}
\end{equation}
where $T_{\mu\nu}$ is the stress-energy tensor. 
This is indeed the case  
provided there exists a four-dimensional dilaton field of the form:
\be 
  \varphi = {\rm ln}r + b'{\rm ln}t  
\label{lineardila}
\ee
which has non-trivial potential 
$V(\varphi)$ such that,
when combined with field-independent contributions 
from the vacuum energy $-\Lambda$,  
one has
\be
\Lambda + V(\varphi) = \frac{2}{r^2}~; 
\label{coconst}
\ee
It is important to check that  
the field $\varphi$ 
satisfies its classical equations of motion in the space-time 
(\ref{fourdime2}):
\begin{equation} 
g^{00} \nabla _0 \partial_0 \varphi + g^{ij}\nabla_i \partial_j 
\varphi = \frac{\delta V(\varphi)}{\delta \varphi}~, i=1, \dots d
\label{scalareq}
\end{equation} 
From (\ref{scalareq}),(\ref{coconst}) one obtains the condition: 
\be
\frac{\delta V(\varphi)}{\delta \varphi}
\left|_{\varphi=\varphi_c}\right. =-\frac{2}{r^2}=-
\left(\Lambda + V(\varphi_c)\right) \label{constr}
\ee
where $\varphi_c$ denotes the configuration (\ref{lineardila}).
From the constraint (\ref{constr}) one then determines $V(\varphi)$ 
as well as the 
contributions to the field-independent part of the vacuum energy $-\Lambda
$:
\be
V(\varphi_c)=\frac{1}{r^2}~,\qquad \Lambda =\frac{1}{r^2} 
\label{vlambda}
\ee
in the $D3$ case considered so far.
In the above, we have ignored the fluctuations 
of the $D3$ brane in the bulk directions. When these are taken into 
account, there may be additional contributions~\cite{emn98,adrian+mavro}
to the 
vacuum and excitation energies on the $D3$ brane, 
which in fact are time-dependent, relaxing to zero asymptotically.

It is interesting to examine whether  
the metric (\ref{fourdime2}) 
and the above analysis for the metric equations can be formally 
extended to $d>3$ bulk (spatial) 
dimensions.
The non-zero components of the Einstein tensor ${\cal E}_{\mu\nu}
=R_{\mu\nu}-\frac{1}{2}g_{\mu\nu}R$ read
in this general case: 
\begin{equation} 
{\cal E}_{00}=0 ~,~{\cal E}_{ij}=-\frac{\delta_{ij}(d-2)}{r^2}-
\frac{x_ix_j}{r^4}~,~i,j=1, \dots d
\label{ddimens}
\end{equation}
We observe from (\ref{fourdime2},\ref{ddimens}) that
the metric 
equations (\ref{einsteinddime})
are satisfied for
the simple case of a free scalar (dilaton) field $\varphi$ 
of the form (\ref{lineardila}), {\it provided }
$d=3$,  
independent of the value of $b'$. 
It seems therefore 
that the restoration of conformal 
invariance in the case of recoiling $D$ particles embedded
in a $Dp$ brane, or equivalently the satisfction of the 
corresponding equations of motion in the Liouville-dressed problem,  
constrains the number of longitudinal 
dimensions on the $Dp$ brane to three.
In other words, {\it only a $D3$ brane can intersect 
with recoiling (fluctuating) $D$ particles}
in a way consistent with the restoration of conformal invariance
in the manner explored here.
 
\section{Energy Conditions and Horizons in Recoil-Induced Space-Times}

It is interesting to look at the energy conditions 
of such space times, which would determine whether
ordinary matter can exist within the horizon region displayed above.
There are various forms of energy 
conditions~\cite{energycond}, which may be expressed as follows: 
\begin{eqnarray} 
{\rm Strong}~~~ &:&~~~ \left(T_{\mu\nu}-\frac{1}{D-2}
g_{\mu\nu}T_\alpha^\alpha\right)\xi^\mu\xi^\nu \ge 0, \nonumber \\
{\rm Dominant}~~~&:& ~~~ T_{\mu\nu}\xi^\mu\eta^\nu \ge 0, \nonumber \\
{\rm Weak}~~~&:& ~~~T_{\mu\nu}\xi^\mu\xi^\nu \ge 0, \nonumber \\
{\rm Weaker}~~~&:& ~~~T_{\mu\nu}\zeta^\mu\zeta^\nu \ge 0.
\label{energyconds}
\end{eqnarray}
where $g_{\mu\nu}$ is the metric and $T_{\mu\nu}$ the 
stress-energy tensor in a $D$-dimensional space-time,
including vacuum-energy contributions,   
$\xi^\mu$ and $\eta^\mu$ are arbitrary future-directed time-like
or null vectors, and $\zeta^\mu$ is an arbitrary null vector. 
The conditions (\ref{energyconds}) have been listed in decreasing
strength, in the sense that
each condition is implied by all its preceding
ones. 

It can be easily seen from Einstein's equations for the metric
(\ref{fourdime2}) 
that inside the horizon $b'^2r^2 \le t^2$ the 
conditions are satisfied, which implies that stable matter can 
exist  {\it inside} such regions of the recoil space-time. 
On the other hand, {\it outside the horizon} 
the recoil-induced metric assumes the form: 
\begin{equation}
ds^{2{(4)}} \simeq 
\left(2 - \frac{b'^2 r^2}{t^2}\right) \left(dt\right)^2 -  
\sum_{i=1}^{3}\left(dx^i\right)^2~,\qquad r^2 > t^2/b'^2
\label{fourdime2b}
\end{equation} 
The induced scalar curvature is easily found to be:
$$R=-4b'^2\left(-3t^2 + b'^2r^2 \right)/\left(-2t^2 + b'^2r^2\right)^2~.$$ 
Notice that there is a {\it curvature} singularity 
at $2t^2 =b'^2r^2$, which is precisely  
the point where there is a signature change in the metric 
(\ref{fourdime2b}).

Notice also that, 
in order to ensure 
a Minkowskian signature in the space-time (\ref{fourdime2b}),
one should impose the restriction  
\begin{equation} 
2 > \frac{b'^2r^2}{t^2} > 
1~; 
\label{signature}
\end{equation} 
Outside this region, the metric becomes {\it Euclidean},
which matches our formal initial construction with a
static Euclidean $D4$ brane embedded in a higher-dimensional bulk 
space-time.  
Notice that, in such a region, one can formally pass to a Minkowskian 
four-dimensional space-time by making a Wick rotation of the Euclidean
time coordinate $X^0$. In this Wick-rotated framework, the
space-time inside the bubbles
retains its Minkowskian signature due to the specific 
form of the metric (\ref{fourdime2}).

The above metric (\ref{fourdime2b}) does not satisfy simple 
Einstein's equations, but this was to be expected,
since the formation of such space-times is not necessarily a 
classical phenomenon~\footnote{On the other hand, the satisfaction of 
conformal invariance conditions on 
a resummed world sheet, as a result 
of Liouville dressing, imply
general `Liouville equations'
in which the $\beta$ functions and the central
charge deficit $Q$ incorporate
higher-genus world-sheet (resummed) effects. Unfortunately,
general expressions for these objects are currently beyond our
calculational reach.}.
Below, we link this fact with the failure 
of the energy conditions in this exterior geometry. 

It can easily be shown that the weaker energy condition
(\ref{energyconds})  
can be satisfied 
for times $t$ such that 
\begin{equation} 
\frac{b'^2r^2}{t^2} \simeq 1+\varepsilon; \qquad \varepsilon \rightarrow 0^+
\label{encon}
\end{equation} 
i.e., on the initial horizon. 
To see this, it suffices to notice that the weaker energy condition
reads in this case:
\be 
     \left(4t^2 - b'^2r^2\right)\left(2 - \frac{b'^2r^2}{t^2}\right)(\zeta^0)^2
\le b'^2\left(\sum_{i=1}^{3} x^i \zeta^i \right)^2 
\label{weak2}
\ee 
where we used the fact that $\zeta^\mu$ is a null vector. 
Choosing $\zeta_1 \ne 0, \zeta_i=0, i=2,3$, it can be shown that 
the right-hand-side of the above inequality can be bounded from above 
by  $$b'^2 r^2 \sum_{i=1}^{3} (\zeta^i)^2=b'^2 r^2 \left(2- 
\frac{b'^2r^2}{t^2}\right)
(\zeta^0)^2~,$$
which, on account of 
the requirement (\ref{weak2}) would imply $\left(2t^2 -b'^2r^2\right)\le 0$. 
This is in contradiction with the range of validity of (\ref{fourdime2b}),
unless one lies on 
the initial 
horizon (\ref{encon}). 
Notice that in this region of space-time there is a smooth matching 
between the interior (\ref{fourdime2}) and the exterior (\ref{fourdime2b}) 
geometries. In such regions 
of space-time, surrounding the recoiling defect, 
matter can exist in a {\it stable form}.

\begin{figure}[htb]
\begin{center}
\epsfxsize=4in
\bigskip
\centerline{\epsffile{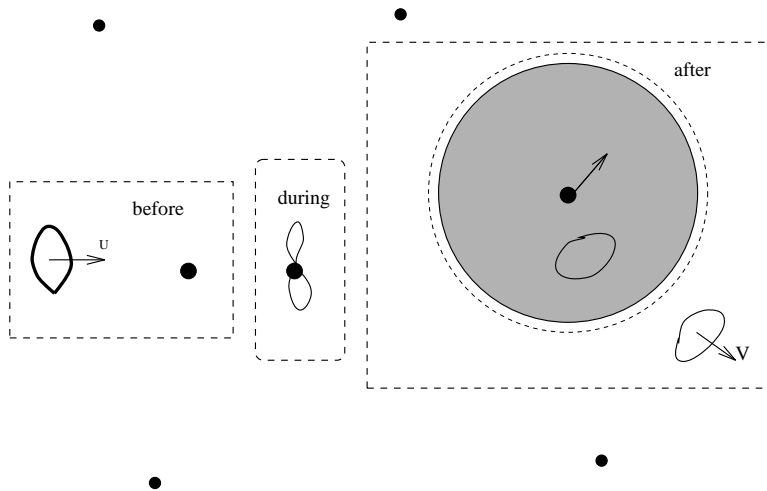}}
\vspace{0.2in}
\caption{\it A schematic representation of scattering in $D$-foam
background. The dashed boxes represent events just before, during
and after the scattering of a closed-string probe on one particular
$D$ brane defect. The scattering results in the formation of
an shaded bubble, expanding as indicated by the dotted line, inside
which matter can be trapped and
there is an energy-dependent refractive index.}
\label{figbubble}
\end{center}
\end{figure}

The above considerations suggest that matter can  be trapped  
{\it inside} such horizon regions around a fluctuating $D$-particle defect. 
This sort of trapping is interesting
for our space-time-foam picture, as it implies that such
{\it microscopic $D$-brane  
horizons act in a similar way as the intuitive description of a
macroscopic black-hole horizon discussed in the Introduction
(\ref{intuition})}, as illustrated in Fig.~\ref{figbubble}.

To reinforce the interpretation that
matter is trapped in the interior of a region
described by the metric (\ref{fourdime2}),
we now show that a matter probe inside the horizon `bubble'
experiences an energy-dependent velocity of light.
First rewrite the metric in a FRW form: 
\begin{equation} 
    ds^2 =e^{2{\rm ln}r} \left(b'^2 dt_{FRW}^2 - \frac{1}{r^2}
\sum_{i=1}^{3} (dx_i)^2 \right)
 \label{scalefactors}
\end{equation} 
where we 
were careful when performing coordinate redefinitions 
{\it not} to absorb in them the factor $b'$, which,  
depends (\ref{varianceb'}) on the energy scale of the matter probe.
We are interested in  matter at various energies
propagating {\it simultaneously} in such a space-time,
and performing a coordinate transformation  
could not absorb an energy-dependent factor such as $b'$.
When we consider the encounter of a matter probe,
such as a photon, with a fluctuating $D$-particle defect, 
the kinetic-energy scale $E_{kin}$ may be identified with the
energy scale $E$ of the matter probe. We recall
that energy conservation has been proven rigorously
in the world-sheet approach to $D$--brane recoil~\cite{mavro+szabo},
and survives the resummation over higher genera.
   
We observe from (\ref{scalefactors}) that the overall scale
factor may be absorbed 
into a redefinition of the spatial part of the dilaton 
(\ref{lineardila}), implying that
stable matter experiences an energy-dependent `light velocity' 
\begin{equation}
  c_{int}(E) = b'c = 2cg_s\left(1 - \frac{285g^2_sE}{18M_Dc^2}\right)^{1/2} 
\label{refrindex}
\end{equation}
in the space-time (\ref{fourdime2}),
where $M_D=M_s/g_s$ is the $D$-particle mass scale. 
The energy-independent factor $2g_s$ may in fact be absorbed
into the normalization of the FRW time coordinate $t_{FRW}$,
thereby making a smooth connection with the velocity of light {\it in
vacuo} in the limiting case of $E/M_Dc^2 \rightarrow 0$.  
{\it It is important to note that, because of the specific form
(\ref{varianceb'}) 
of the variance $b'$, the resulting effective velocity (\ref{refrindex})  
in the interior of the bubble is subluminal~\cite{ellis99}}. 
On the other hand, we see from (\ref{fourdime2b}) that
matter propagates at the normal {\it in vacuo} light velocity 
$c$ in the exterior part of the geometry.

If one considers pulses containing many photons of different
energies~\cite{sarkar,efmmn}, then 
the various photons will experience, as a 
result of the dynamical formation of horizons, changes in 
their mean effective velocities corresponding on average to a
{\it refractive index} $\Delta c(E)$, where
the effective light velocity:
\begin{equation}
c(E)=c\left(1 - \xi \frac{g_sE}{M_sc^2} \right).
\label{refind}\end{equation} 
Here $\xi$ is a quantity that depends on the actual details of the 
scenario for quantum space-time foam, in particular on the density of the
$D$-brane defects in space. 
In a dilute-gas approximation, $\xi$ might plausibly be assumed to be of
order one, as can be seen as follows.
Consider a path $L$ of a photon, which encounters ${\cal N}$ 
fluctuating $D$-particle defects. Each defect creates a bubble
which is expected to be close to the Planckian size $\ell_s$, for any 
reasonable model of space-time foam.
Inside each bubble, the photon propagates with velocity (\ref{refrindex}),
whereas outside it propagates with the velocity of light in vacuo $c$.
The total time of flight for this probe will therefore be given by: 
\begin{equation} 
t_{total} = \frac{L-{\cal N}\ell_s}{c} + {\cal N} \frac{\ell_s}{c}
\left(1 - g_s^2\frac{285}{18}\frac{E}{M_Dc^2}\right)^{-1/2}
\label{flighttime}
\end{equation}
In a `dilute gas approximation' for the description of space time 
foam, it is natural to assume that a photon encounters, on average,  
${\cal O}(1)$ $D$ particle
defect in each Planckian length $\ell_s$, so that
${\cal N} \sim \xi L/\ell_s$, where $\xi \le 1$.
From (\ref{flighttime}), then, one obtains a delay in the arrival time 
of a photon of order
\begin{equation} 
\Delta t \sim \xi g_s^2 \frac{285}{36}\frac{L~E}{M_Dc^3} + \dots,
\label{finaldelay}
\end{equation}
corresponding to the effective velocity (\ref{refind})~\footnote{In
conventional 
string theory, $g_s^2/2\pi \sim 1/20$, and the overall numerical factor
in (\ref{finaldelay}) is of order $4.4~\xi$. However, $g_s$ 
should rather be considered an arbitrary parameter of the model, which may
then 
be constrained by phenomenological observations~\cite{efmmn} through 
limits on (\ref{finaldelay}).}.

\section{Breathing Horizons in Liouville String Theory}

The tendency of the horizon (\ref{encon}) to expand is a classical 
feature. Upon quantization, which corresponds
in our picture to a proper resummation
over world-sheet topologies, one expects a phenomenon similar to Hawking 
radiation. Such a phenomenon would decelerate and stop the expansion,
leading eventually to the shrinking of the horizon. 
This would be a dynamical picture 
of space-time foam, which unfortunately at present is not fully available,
given that at microscopic distances the world-sheet perturbative 
analysis breaks down. However, we believe 
that this picture is quite plausible, and
we can support these considerations formally
by recalling that time $t$ is the 
Liouville field in our formalism. 

\begin{figure}[htb]
\begin{center} 
\epsfig{figure=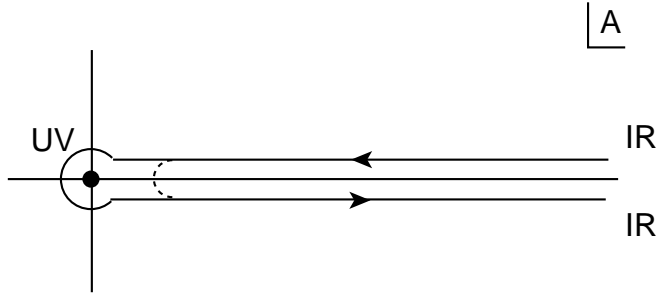}
\caption{{\it Contour
of integration appearing in the analytically-continued
(regularized) version of world-sheet Liouville string correlators.
The quantity $A$ denotes the (complex) world-sheet area. 
This is known in the literature as the Saalschutz contour, and
has been used in
conventional quantum field theory to relate dimensional
regularization to the Bogoliubov-Parasiuk-Hepp-Zimmermann
renormalization method. Upon the identification of the 
Liouville field with target time~\cite{emn98}, this curve
resembles closed-time paths in non-equilibrium field
theories~\cite{ctp}.}} 
\label{bounce}
\end{center}
\end{figure}

\begin{figure}[htb]
\begin{center}
\epsfxsize=3in
\bigskip
\centerline{\epsffile{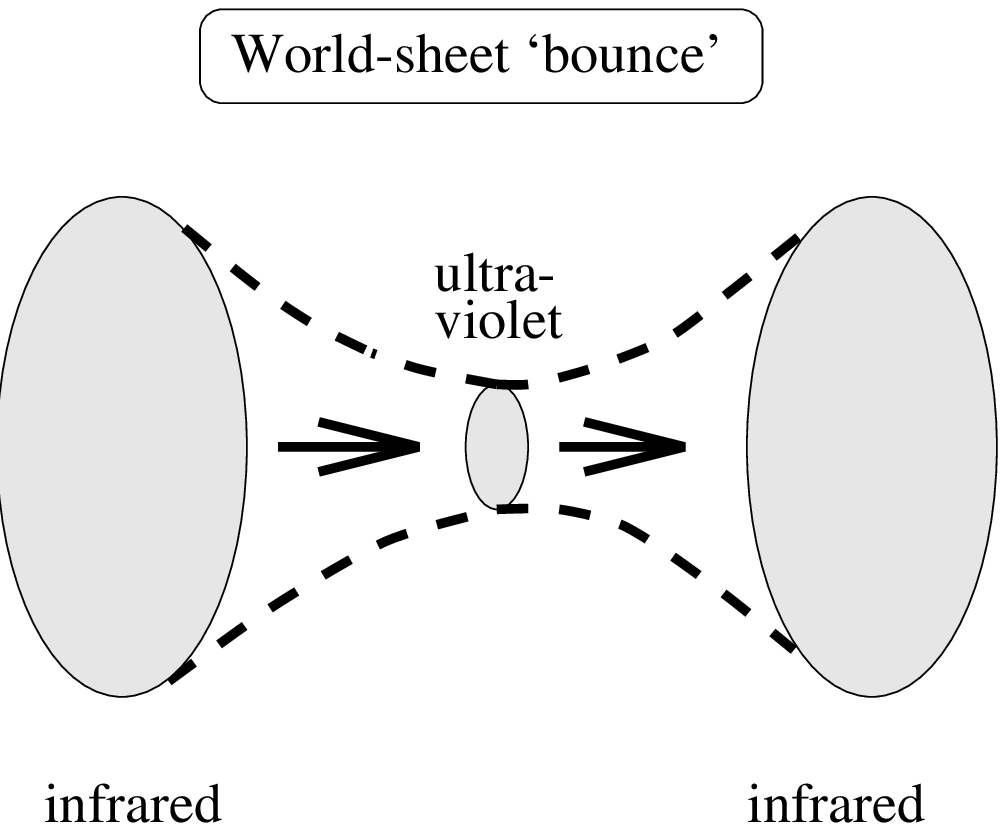}}
\caption{{\it The evolution of the world sheet of the Liouville string,
when the Liouville mode lies on the (dashed) curve of Fig.~\ref{bounce}. 
Upon interpreting the (zero-mode of the) 
Liouville field as time, the above `bounce' evolution
suggests a `breathing mode' for the Liouville Universe, which from our 
point of view represents a formal description of a space-time 
foam `bubble'.}} 
\label{bounce2}\end{center}
\end{figure}

As we have pointed out previously, the dynamics of the
Liouville field exhibits a `bounce' behaviour, when considered
from a world-sheet view point~\cite{kogan,emn98}, as illustrated 
in Figs.~\ref{bounce} and \ref{bounce2}. 
This is a general feature of non-critical strings,
whenever the Liouville field is viewed as a local 
renormalization-group scale
of the world sheet. As emphasized in~\cite{ellis98}, the bounce picture 
is necessitated by the decomposition of the Liouville world-sheet
correlators
as ordinary scattering-matrix elements in target space. Specifically,
these correlators diverge with the world-sheet area scale $A$ in the 
infrared limit $A \rightarrow 0^+$. One may regularize such divegences
by defining the  world-sheet path-integral over the Liouville mode on 
the analytically-continued curve illustrated in Fig.~\ref{bounce}. 
When one inteprets the (zero mode of the) Liouville field $\phi$ 
as time~\cite{emn98}: $t \propto {\rm log}A$, therefore, 
the contour of Fig.~\ref{bounce} 
represents evolution in time, as seen in Fig.~\ref{bounce2},
in both directions between fixed points of the 
world-sheet renormalization group: $ {\rm Infrared} ~ {\rm fixed} ~ 
{\rm point}  \rightarrow {\rm  Ultraviolet} ~ {\rm fixed}
~{\rm point} \rightarrow
 {\rm Infrared} ~ {\rm fixed} ~ {\rm point}$.

When one integrates over the Saalschultz contour in
Fig.~\ref{bounce}, the integration
around the simple pole at $A=0$ yields an imaginary part~\cite{kogan,emn98},
associated with the instability of the Liouville vacuum. We note, on the  
other hand, that the integral around the dashed contour
shown in Fig.~\ref{bounce}, 
which does not encircle the pole at $A=0$, is well defined.
This can be intepreted as a well-defined $\$$-matrix element,
which is not, however, factorisable into a product of 
$S-$ and $S^\dagger -$matrix elements, due to the 
$t$ dependence acquired after the identification 
$t=-{\rm log}A$~\footnote{This formalism is similar to the 
Closed-Time-Path (CTP) formalism used in non-equilibrium 
quantum field theories~\cite{ctp}.}. The absence of
factorization is linked to the evolution (\ref{intuition}) from
a pure state $\vert A, B \rangle$ to a mixed density matrix
(\ref{mixedrho}), which cannot be described by a conventional $S$ matrix.

In our approach, the logarithmic algebra of the recoil 
operators forces 
the regularizing parameter $\epsilon$ (\ref{defeps})
to be identified with the logarithm of the
world-sheet area scale 
$A=\left|L/a \right|^2$, and hence with the target time. 
In the bounce picture outlined above, there will be a `breathing mode'
in the recoil-induced space-time, {\it characterized by
two directions of time}, 
corresponding to the processes of expansion, stasis and 
shrinking of the horizon in the recoil-induced space-time
(\ref{fourdime2}), all
within a few Planckian times. 
This is the Liouville-string
description of Hawking radiation.

\section{Outlook} 

We have discussed in this article a microscopic mechanism for the 
dynamical formation of horizons by the collisions of closed-string
particle `probes' with recoiling $D$--particle
defects embedded in a $p$-dimensional 
space time, which may in turn be viewed as
a $Dp$ brane domain wall in a higher-dimensional target space. 
As we have argued before,
the correct incorporation of recoil effects, which are unavoidable 
in any quantum theory of gravity that reproduces the 
conceptual framework of general relativity in the classical limit, 
necessitates a Liouville string approach in the context
of a (perturbative) world-sheet framework. 

The most important result of our approach in this paper is the
demonstration of the dynamical formation of breathing horizons, 
which follows directly from the 
restoration of conformal invariance by means
of Liouville dressing. The horizon
regions were discovered using the positive-energy
theorems for the recoil-induced space-time. We have been able to show
that such regions form bubbles with a non-trivial 
refractive index, with light propagation that
is always subluminal~\cite{ellis99}, because of the specific properties
of the space-time induced by our treatment of $D$--particle recoil.
The breathing nature of the horizons, which follows 
from specific properties of the Liouville dynamics, 
is the best candidate we have in this framework for 
a quantum space-time foam, generalizing
appropriately the
Hawking radiation of conventional macroscopic black holes to 
the microscopic $D$--brane case. 

The non-trivial optical properties induced by
the propagation of light in such a fluctuating space-time
may be subject to experimental verification 
in the foreseeable future, and are alreday
constrained by existing data~\cite{sarkar,efmmn}. 
The fact that the refractive index in the bubbles of
space-time foam is subliminal implies the absence 
of birefringence in light propagation, which is, however,
possible in other approaches to space-time foam~\cite{pullin}. 

One curiosity of our analysis has been that
the requirement of restoring conformal 
invariance by means of the Liouville field, which in 
our approach is identified with the target time, 
is quite restrictive. It implies
within this approach that {\it only $D3$} branes can intersect
consistently with fluctuating $D$ particles. 
The result is surprising, as it seems to provide
a mathematical reason for the fact that we live in 
four dimensions only. However, the model is oversimplified,
in the sense that only dilaton and graviton fields have been 
considered so far in modelling the dynamics of the distortion 
of the space-time due to the fluctuating defect. 
This may in some sense be analogous to the way the critical dimension
of conventional string theories was revealed in the 
context of a $\sigma$-model approach, when only the conformal anomaly
contributions in a flat target space were considered.
However, there have been many independent
confirmations of the critical dimension in this `traditional' string case, 
coming, e.g., from the the
no-ghost theorems and the closure of the Lorentz algebra.
In the present case, we currently lack further support 
of our result from an independent calculation, but
we consider it as worthy of further investigation. 

Leaving aside this issue of the critical dimension,
our theoretical model is admittedly crude, and should by no means
considered as complete. However, we believe that it provides 
a concrete example how space-time foam
might arise in the context of modern string/$D$--brane theory. 
Certainly much more work, both theoretical and `phenomenological'
is necessary before even tentative conclusions are reached on 
these important matters. But it is our firm belief that 
the model presented here contains the seeds
for an eventual understanding of many 
important issues associated with the space-time foam 
structure of quantum gravity, and for this reason it 
deserves further and more detailed studies
before it can be excluded.

\section*{Acknowledgements}

The work of N.E.M. is partially supported by PPARC (UK)
through an Advanced Fellowship. 
That of D.V.N. is partially supported by DOE grant 
DE-F-G03-95-ER-40917.
N.E.M. and D.V.N. also thank
H. Hofer for his interest and support.

\end{document}